\documentclass[12pt,a4paper]{article}
\usepackage{amssymb}

\def \AO {{\widehat A}}

\def \CO {{\widehat C}}

\def \PO {{\widehat P}}
\def \pO {{\hat p}}
\def \qO {{\hat q}}
\def \RO {{\widehat \rho}}
\def \SO {{\widehat S}}
\def \DO {{\widehat W}(q,p)}
\def \WO {{\widehat W}}

\def \be {\begin{equation}}
\def \ee {\end{equation}}
\def \bea {\begin{eqnarray}}
\def \eea {\end{eqnarray}}
\def \mea {\nonumber\\}

\begin{document}
\title{Non-Positivity of the Wigner Function   and Bounds\\ on Associated Integrals}
\author{
A.J. Bracken$^{1,a}$, D. Ellinas$^{2,b}$ and J.G. Wood$^{1,c}$ 
\\
$^1$ Centre for Mathematical Physics, Department of Mathematics \\
University of Queensland, Brisbane 4072, Australia\\
$^2$ Department of Sciences, Section of Mathematics\\
Technical University of Crete, GR-731 00 Chania, Crete, Greece
\\\qquad\\
Paper presented at the Wigner Centennial Conference, \\P\'ecs, Hungary, July, 2002\\
(to appear in the Proceedings)}
\date{}

\maketitle
\begin{abstract}{The Wigner function shares several properties with classical distribution functions on
phase space, but is not positive-definite.  The integral of the Wigner function over a given
region of
phase space can therefore lie outside the interval $[0,1]$.
The problem of finding best-possible upper and lower bounds for a given region is the problem of
finding the greatest and least eigenvalues of an associated Hermitian operator.
Exactly solvable examples are described, and possible extensions are indicated.  }
\end{abstract}

\section{Introduction}\label{intro}
Since its introduction \cite{wigner}
in 1932, the Wigner distribution function has proved an important
tool  in many areas of
quantum physics and chemistry.
Recently, in quantum tomography \cite{leonhardt},
the Wigner function has been
reconstructed from experimentally
measured probabilities associated with a quantum system in a definite state.
Given a finite amount of data, it is clear that such a
reconstruction must inevitably be limited to a finite region of phase space.
And here there arises a problem peculiar to quantum tomography.  In the classical case, if the
probability density has been reconstructed on
a finite region, and if its integral over that region,  the
`probability mass' on the region, is close to 1, then it is certain that the density on the
remainder of phase space is small, in the sense that the probability mass there must be small.
But in the quantum case, because
the Wigner function
takes negative as well as positive values in general,
the `quasiprobability mass' on regions of phase space outside one particular region may
be large positive or negative, even when the
quasiprobability mass on that particular  region  is known to be close to 1.
This is the primary motivation for our study, which aims to provide best possible upper and lower
bounds on the integral of the Wigner function over any given region $S$ of phase space.

In previous work \cite{bracken}, it has been shown that this is equivalent to finding the
greatest and least eigenvalues
of a corresponding  Hermitian operator ${\widehat S}$, which we refer to as the
`region
operator' for $S$.
It is an operator analogue of the classical characteristic
function ${\delta}_S$, which has the value $1$ for points inside $S$, and $0$ for points
outside.
Here we describe some exact results and the basis on which they can be obtained, and
indicate some directions of possible generalization.

\section{Mathematical background and results}

For a system with one degree of freedom,
the invertible mapping from functions $A(q,p)$ on the phase plane $\Gamma$
to linear operators ${\widehat A}$
on Hilbert space is
accomplished with the help of  Weyl's kernel operator
\cite{weyl},
\begin{eqnarray}
\DO&=&\frac{1}{(2\pi\hbar)^2}\int_{\Gamma} e^{i[p' (q-\qO)-q' (p-\pO)]/\hbar}
\, dq'\, dp'\,,
\mea
\AO=\int_{\Gamma} A(q,p) \DO \,dq\,dp\,,&&\quad A(q,p)=2\pi\hbar\,\rm{Tr}(\AO\DO)\,.
\label{mapdef}
\eea
Here Tr denotes the trace, and $\qO$, $\pO$ are canonical operators.
The generalizations to many degrees of
freedom are obvious. From this point we work with dimensionless variables, in effect setting
$\hbar=1$.

As  special cases, we have the Wigner function $W_{\rho}(q,p)$ coresponding to
the operator $\RO/(2\pi)$,
where $\RO$ is the density operator defining the state of the quantum system;
and the region operator $\SO$,
corresponding to
the characteristic function $\delta_S(q,p)$ of
a given 2-dimensional region $S\in\Gamma$\,:
\be
W_{\rho}(q,p)=\rm{Tr} (\DO\RO)\,,\quad \SO=\int_{\Gamma}
\delta_S(q,p) \DO\,dq\,dp=\int_S \DO\,dq\,dp\,.
\label{chishdef}
\ee
In the last step we have used the defining property of the characteristic function.

Given a smooth contour $C\in\Gamma$, it is natural to define also
an associated `contour operator'
$\CO$ by
\be
\CO=
\int_{\Gamma} \DO\delta_C(q,p)\,dq\,dp\,.
\label{Cdef1}
\ee
Here $\delta_C(q,p)$ is the `characteristic (generalised) function' for the contour,
defined by the property that, for every smooth function $F$ on $\Gamma$,
\be
\int_\Gamma F(q,p)\delta_C(q,p)\,dq\,dp
=\int_C F\,dl\,,
\label{deltadef}
\ee
where $dl$ the element of length on $C$.
Then we can write
\be
\CO=\int_C \WO\,dl\,.
\label{Cdef2}
\ee
For example, if the contour is defined by the graph of a simple function, such as $p=F(q)$,
$a<q<b$, then $\delta_C(q,p)=\delta(p-F(q))$ for $a<q<b$, and $\delta_C (q,p)=0$ elsewhere.
Going one dimension further down, we can define the `point operator'
\be
\PO=\int_{\Gamma} \delta_P(q,p)\DO\,dq\,dp=\WO(q_0,p_0)\,.
\label{pointopdef}
\ee
corresponding to the point $(q_0,p_0)\in\Gamma$.    The characteristic function for
the point, $\delta_P(q,p)=\delta(q-q_0)\delta(p-p_0)$,
is more singular than that for a contour $C$, which is more
singular than that for a 2-dimensional region $S$ in the plane.  Clearly these notions can be
generalized to regions $R$ of dimensions $2f$, $2f-1$, $\dots$, $0$ in the $2f$-dimensional phase
space of a system with $f$ degrees of freedom, with the introduction of corresponding characteristic
functions $\delta_R$ of increasing singularity, and corresponding region
operators ${\widehat R}$.

Critical to what follows is Wigner's observation \cite{wigner}
that, when the quantum system is in the state $\RO$,
the mean value of any operator (such as ${\widehat R}$)
is given in terms of the corresponding phase space function (in this case $\delta_R$) and
the Wigner function, by
\be
\langle {\widehat R}\rangle =\int_{\Gamma} \delta_R W_{\rho} \,d\Gamma=\int_R W_{\rho} \,dV\,,
\label{expec}
\ee
where $dV$ is the volume element in $R$.  Since the expectation value
$\langle{\widehat R}\rangle$ must lie between the greatest and least eigenvalues of ${\widehat R}$
(or more generally, between the best possible upper and lower 
bounds on its spectrum), the same must be true of the last integral in
(\ref{expec}). Moreover, the bounds are attained (or more generally, approached arbitrarily
closely) if $\RO$ is a pure state corresponding to the associated (generalized) eigenvector of
${\widehat R}$.  In short, if we can solve the eigenvalue problem for ${\widehat R}$, we have best
possible bounds on the last integral in (\ref{expec}), and we know when  and how those bounds are
attained.

The circular disk of radius $a$,
centred on the origin in the phase plane is invariant under canonical
transformations generated by
the classical harmonic oscillator Hamiltonian $H=p^2+q^2$, and it follows that
the corresponding region operator ${\widehat S_a}$ commutes with the quantum oscillator
Hamiltonian operator,
and has the same eigenvectors, namely the number states $|n\rangle$, $n=0,\,1,\,\dots$. This is
significant for quantum tomography, where such states are amongst the easiest to measure
\cite{leonhardt}. The eigenvalue of ${\widehat S_a}$ on $|n\rangle$ is \cite{bracken}
\be
\lambda_n(a)=(-1)^n\int_{0}^{a^2}L_n(2u)\,e^{-u}\,du\,,
\label{diskevals}
\ee
where $L_n$ is the Laguerre polynomial.
For a given $a$, it is then possible
to see which eigenvalue 
is smallest (say $\lambda_N(a)$), and which is  largest (always $\lambda_0(a)$), and
hence what are the lower and upper
bounds on the integral of the Wigner function over the disk.  These bounds are
attained when the system is in the pure state $|N\rangle$ or $|0\rangle$, respectively.
The eigenvalues of the contour operator ${\widehat C}_a$ corresponding to the boundary of the
circular disk are, again on the number states, $\mu_n(a)=d\,\lambda_n(a)/d\,a$.
There is a remarkable Lie-algebraic  
structure
\cite{ellinas}
associated with these spectra.

For the contour operator ${\widehat C}_L$ corresponding to
a straight line segment of length $L$ in the phase plane, say the segment along the $p$-axis
from $p=-L/2$ to $p=L/2$,
we find \cite{wood} a continuous spectrum and generalized eigenfunctions:
\be
 \mu(k)=\pm \frac{1}{k\pi}\sin (kL)\,,\,\,(0\leq k<\infty)\,;\,\,\,\,
\varphi_k(x)=\frac{1}{\sqrt{\pi}}\,\bigg\{{\cos(kx) \atop \sin(kx)}\,.
\label{segmentevals}
\ee
This segment is invariant under canonical transformations generated by the free-particle
Hamiltonian $H=p^2$, and also under the parity transformation which sends $q$, $p$ to $-q$,
$-p$, hence the appearance of the common eigenfunctions of the quantum free-particle Hamiltonian
operator
and parity operator in (\ref{segmentevals}).

From these simple results for disks, circles and straight segments we can build up results for
more complicated regions in phase spaces of higher dimensions.
For example, in the 4-dimensional phase space of a 2-dimensional quantum system, we can find
best possible upper and lower bounds on the integral of the Wigner function on the rectangle
$-L_1/2<q_1<L_1/2$, $-L_2<q_2<L_2/2$ within
the `Poincar\'e
section'  $p_1=p_2=0$; or on the cylinder $q_1^2+p_1^2\leq a^2$, $-L/2<p_2<L/2$, $q_2=0$; or on
its surface, and so on.

Investigations continue towards a ``spectral theory" of quasiprobability mass
for bosonic systems, and 
into analogous results for spin and other
finite dimensional quantum systems, with expected 
applications in
quantum tomography. 

\section*{Notes}
\noindent
Email:

\noindent
a. ajb@maths.uq.edu.au \,
b. ellinas@science.tuc.gr \,
c. jgw@maths.uq.edu.au

\vfill\eject
\end{document}